\documentstyle[11pt]{article}
\begin{document}

\hfill PREPRINT INR-0952/97

\hfill August 1997
\vskip 2cm

\begin{center}
{\Large
\bf Analytical continuation and resummed perturbation theory}

\vskip 2cm

{\large
\bf A.A. Pivovarov}

{\it Institute for Nuclear  
Research of the Russian Academy of Sciences, 
Moscow 117312, Russia}

\vskip 3cm
{\bf Abstract}
\end{center}

A pattern of 
partial resummation of perturbation theory series
inspired by analytical continuation is
discussed
for some physical observables.

\vskip 8cm
\begin{center}
Talk given at QCD 97 Conference, Montpellier, France 

4 -- 9 July 1997
\end{center}

\newpage 

\section{INTRODUCTION}
At low energy $\mu\sim m_\tau$
an expansion parameter of perturbation 
theory the strong coupling constant 
$\alpha_s = \alpha_s(m_\tau)\simeq 0.35$ 
is rather large (e.g. \cite{altrev}) 
and, therefore,
for a generic QCD observable $\sigma$
(without Born term)
\begin{equation}
\sigma \sim \alpha_s(1+\sigma_1\alpha_s+\sigma_2\alpha_s^2
+\ldots)
\label{first}
\end{equation}
the series of perturbation theory approximation
converges badly the last term being about 10\% of the leading one.

To improve predictions that is required by experimental data 
for some observables at present,
higher order corrections have to be included. 
This, however, is technically
difficult due to the necessity of computing many-loop integrals
representing 
Feynman diagrams in high orders of perturbation theory. 
There is a little hope to obtain next terms in (\ref{first})
(beyond three-four loops) for many processes in realistic models.

It should be stressed that phenomenologically (leaving apart general
considerations of possible power corrections
that are necessary for two point correlators and lead to resonances) 
there is no much intrinsic reason to go beyond perturbation theory:
wild asymptotic behavior is not seen yet. Also the freedom of the choice
of the renormalization schemes allows one to render the series convergent
for a given observable (or some set of observables) at least
in an heuristic sense that further known terms decrease \cite{crit}.
There is no strict indication 
that perturbation theory is broken though 
the accuracy it can provide in a number of cases is not sufficient 
for confronting predictions  with experimental data.
Thus, at the level of phenomenology 
just needs of precision require
an improvement of perturbation theory
predictions
and because further terms are not available going
beyond perturbation theory in different ways
is now widely discussed \cite{Nordita}.

Before adding genuine nonperturbative terms 
which is not obvious in cases when Wilson operator product expansion
is not directly applicable one tries to go beyond the
finite order perturbation theory by 
a resummation of a particular subset of terms that
can be explicitly generated. The simplest one is due to running of the
coupling constant.

This resummation is ambiguous to a great extent in particular
it can change the analytic properties that exist in any
finite order of perturbation theory and are established on a general
ground of quantum field theory and even can make them wrong 
\cite{resum}, i.e. resummed quantities can not satisfy some general
requirements that leads to necessity of interpretation of the results.
The ambiguities that are 
produced by the resummation and the change of analytic properties 
are analyzed in some details.

\section{AMBIGUITY FOR THE $\tau$ LEPTON WIDTH IN MS SCHEME}
The $\tau$ lepton
width became a real laboratory for investigation of 
properties and numerical validity of low energy perturbation theory
\cite{BraNarPic}.

The spectral density $R(s)$ for a two point correlator 
\[
\Pi(x)=\langle 0|Tj(x)j(0)|0\rangle
\]
where $j(x)$ is a weak charged current
of light quarks generates Adler's function 
\begin{equation}
D(Q^2)=Q^2\int_0^\infty{R(s)ds\over (s+Q^2)^2},~Q^2=-q^2
\label{adl}
\end{equation}
that can be calculated in Euclidean domain in terms of 
perturbation theory series in the coupling constant $\alpha_s(\mu)$
\[
D(Q^2)=\alpha(\mu)+\alpha(\mu)^2(\beta_0 \ln{\mu^2\over Q^2}+c)+\ldots
\]
From the last expression the spectral density can be found in any
finite order of perturbation theory in the form  
\begin{equation}
R(s)=\alpha(\mu)+\alpha(\mu)^2(\beta_0 \ln{\mu^2\over s}+c)+\ldots
\label{rs}
\end{equation}
The $\tau$ lepton decay width are given by 
\begin{equation}
r_\tau=\int_0^{m_\tau^2}R(s)W(s/{m_\tau^2})ds/{m_\tau^2}, 
\label{tauwidth}
\end{equation}
where
\[
W(x)=(1-x)^2(1+2x).
\]
In a
finite order of perturbation theory the expression 
for the $\tau$ lepton width has the form  
\[
r_\tau=\alpha(\mu)+\alpha(\mu)^2(\beta_0 \ln{\mu^2\over
m_\tau^2}+\tilde c)+\ldots
\]
Above formulas are given for normalization only 
and $\beta_0$ is the first
coefficient of the $\beta$ function in QCD.
Within the formal perturbation theory in the strong coupling constant 
$\alpha_s$ every term ($ln^n{\mu^2\over Q^2}$)
has correct analytic properties in $q^2$: a cut along the positive semiaxes.
Using formulas (\ref{adl}) and (\ref{rs}) with known renormalization
group properties of $R(s)$ and $r_\tau$ ($\mu$-independence) 
one can go beyond the predictions of finite order
perturbation theory and
perform a partial resummation in many different ways.
First it can be done directly on the cut \cite{KraPivRen}. Because 
the spectral density itself is nonintegrable at low energy after using
the renormalization group improvement
on the cut 
($\mu^2=s$)
\[
R(s)\sim\alpha_s(s)={1\over \beta_0 \ln (s/\Lambda^2)}
\]
one can  
take the discontinuity after RG summation in Euclidean domain
to obtain (in the leading order)
\begin{equation}
R(s)= {1\over \pi}\arctan \pi \alpha(s)=\alpha(s)
-{\pi^2\over 3}\alpha(s)^3
+\ldots
\label{arctan}
\end{equation}
Now the integral in (\ref{tauwidth}) can be done that provides an
improvement of PT prediction through resummation on the cut.
This technique works up to third order of PT that is available now,
i.e. commutation of RG summation and taking the discontinuity
makes the integral in (\ref{tauwidth}) regular. 

At third order of perturbation theory, however, there appears
another way of definition of the quantity in question (\ref{tauwidth})
that is
connected with the change of renormalization scheme that reduces to
a redefinition of
the coupling constant. In the effective charge scheme \cite{Grunberg}
an effective 
$\beta_\tau(a_\tau)$ has a zero that leads to an IR fixed point \cite{Ste}
and integrals (\ref{tauwidth})  can be explicitly done in this scheme
\cite{KraPivRen}. This possibility however depends crucially on the
order of PT and is absent in the second order. It is unknown whether
it persists in the fourth order. So for this technique the natural
requirement that the method of resummation is stable in every order of
PT is not fulfilled.

There are 
also other possibilities of resummation, for instance, different kinds
of optimization
\cite{Steinv,OPT}.

Another recipe (more perturbative because it is formulated in the
complex plane and not on the physical cut) 
is to use the analytic properties
given by (\ref{adl}) and define \cite{tau} 

\[
\int R(s)ds = {1\over 2\pi i}\int_C \Pi(z)dz.
\]
Cauchy theorem requires no singularities inside the contour 
so if 
\[
D(z)\simeq D(z)^{PT}\sim \alpha(z)={1\over \beta_0 \ln (-z/\Lambda^2)}
\]
then
$D(z)^{PT}$ has wrong analytic properties 
and there is a difference 
(nonperturbative) which is proportional to 
$\Lambda = m_\tau \exp(-2\pi/9\alpha_s( m_\tau))$ 
with the result of direct summation on the cut.

One should stress that the modified minimal subtraction scheme is
always used for the definition of the charge and the change 
of an observable within PT is
formally of higher order in the coupling constant. The real problem 
is that this difference is large enough to be caught
by experiment. Then the theoretical predictions can differ by the
amount that depends on the procedure used in computation
and is not negligible.
It is unclear how to single out the best numerical value.
There is here even more ambiguity than the simple freedom in the
choice of the renormalization scheme.

Thus,
$\alpha_{MS}(m_\tau)$ depends rather strongly 
on the resummation procedure
that should be explicitly explained when precise comparison of
different predictions is made. 

Having this ambiguity in mind and noting that 
$\alpha_{\overline{\rm MS}}(m_\tau)$ by
itself can not be measured because it is unphysical quantity we next
consider more strict test of pQCD that involves only observables 
and therefore is 
free of the renormalization scheme ambiguity. Still an ambiguity due
to resummation contrary to the finite order analysis \cite{ptest} 
is present.

\section{A TEST OF pQCD FOR DIRECTLY MEASURED\\ QUANTITIES}
For the analysis 
the moments of $e^+e^-$ annihilation and $r_\tau$
are chosen because \cite{ptest,adler}
\begin{itemize}
\item
these observables 
are generated by the same Green's function in pQCD ($m_q=0$)
that reduces unknown possible nonperturbative effects
\item
the integration scale for the moments can be adjusted 
in such a way to avoid the renormalization group evolution
that would require the use of the $\beta$ function and introduce 
further uncertainties
\item 
both moments and $r_\tau$
can be directly measured with high precision
that allows to pin down the theoretical difference that is 
parametrically of the next 
(fifth) order in the coupling constant and is fairy small.
\end{itemize}
Notations for further analysis are as follows \cite{adler}.
The whole spectral density 
\[
R(s)=2(1+{4\over 9}r(s))
\]
is defined through the reduced one 
\[
r(s)={9\over 4\pi} \alpha+\ldots
\]
that determines the reduced Adler's function
\[
\tilde d(Q^2)=Q^2\int_0^\infty{r(s)ds\over (s+Q^2)^2}.
\]
\[
=
{\alpha\over \pi}+k_1\left({\alpha\over \pi}\right)^2
+k_2\left({\alpha\over \pi}\right)^3
+\tilde k_3\left({\alpha\over \pi}\right)^4+\ldots
\]
The moments of $e^+e^-$ annihilation rate are defined by
\[
r_n=(n+1)\int_0^{m_\tau^2}{ds\over m_\tau^2}
\left({s\over m_\tau^2}\right)^n r(s)
\]
\[
r_\tau=2 r_0 - 2 r_2 + r_3.
\]

Coefficients $k_i$ summarize all information from perturbation theory
and are the only ingredient for testing the theory. We factor out all
known nonperturbative corrections due to condensates.  
The technique of resummation based on integration along the contour
in the complex plane is adopted \cite{tau}.
Because $r_\tau$  and $r_n$ are 
renormalization group invariant it is convenient to use
renormalization group invariant approach from the the very beginning.
Introduce 
\[
d_\tau=d(m_\tau^2)=m_\tau^2\int_0^\infty{r(s)ds\over (s+m_\tau^2)^2}
\]
for which the RG equation is 
\[
z{d\over dz} d = - d^2(1+\rho_1 d+\rho_2 d^2+\rho_3 d^3+\ldots)
\]
where $\rho_i$ are renormalization scheme invariants,
$\rho_1=0.79$, $\rho_2=1.035$, $\rho_3=2 k_3-2.97953$
($\overline{\rm MS}$ parameterization) 
with recently computed coefficient 
of the $\beta$ function \cite{Ritbergen}.

Moments are defined as integrals along the contour in the complex
plane
and therefore unphysical singularity
is included.
If 
\[
p(z): ~ -z{d\over dz}p(z)=d(z)
\]
then 
\[
r_n={n+1\over 2\pi i}\int_{|x|=1}x^n p(m_\tau^2x)dx.
\]

The machinery of computing consists now in finding functions $f_n(.)$
and $g(.)$ and then inverting the function $g(.)$ to represent moments
through the only input parameter $r_\tau$
\[
r_n=f_n(d_\tau), \quad r_\tau=g(d_\tau), \quad d_\tau=g^{-1}(r_\tau),
\]
\[
r_n=f_n[g^{-1}(r_\tau)]=(f_n\otimes g^{-1})(r_\tau).
\]
Analytic properties of the function $f_n\otimes g^{-1}(.)$ in $r_\tau$
determine a radius of convergence for our series \cite{ptest}.
We could not determine these properties completely.
However the simpler question -- analytic 
properties of $f_n(x)$ with respect
to $x$ -- can be answered completely (in 
$\overline{\rm MS}$ scheme without $k_3$).
To the leading order 
the explicit formula is 
\[
f_0(x)={1\over 2\pi x}\int_{-\pi}^\pi{e^{i\phi}d\phi\over 1+i x \phi}=
1+2 x+\ldots
\]
that gives 
$x<1/\pi$ or $\alpha < \frac{4}{9}\frac{1}{\pi}$ \cite{LeDiberderPich}.

Generalization to higher orders is straightforward, one has to find 
a singularity of the solution of 
the renormalization group equation \cite{resum}
\[
z{d\over dz} d(z) = \beta(d(z)), \quad d(m_\tau^2)=d_\tau.
\]
The result is
$\alpha^{(1)} < \frac{4}{9}0.744$,  
$\alpha^{(2)} < \frac{4}{9}0.697$,
$\alpha^{(3)} < \frac{4}{9}0.674$ \cite{resum}.
 
The actual value of $\alpha^{(3)} = 0.3540$ 
obtained from 
$r_\tau^{exp}=0.487\pm0.011$ \cite{Pich}
lies outside convergence regions.
The behavior of the decay rate in higher orders of perturbation theory
is presented in Table \ref{tab1} \cite{resum}.
\begin{table}\begin{center}
\begin{tabular}{|r|l||r|l||r|l|}\hline
 $n$&$r_\tau$&$n$&$r_\tau$&$n$&$r_\tau$\\\hline
 $1$&$0.2535$&$5$&$0.4993$&$9$&$0.7400$\\
 $2$&$0.4021$&$6$&$0.4622$&$10$&$1.5577$\\
 $3$&$0.4870$&$7$&$0.4402$&$11$&$4.059$\\
 $4$&$0.5153$&$8$&$0.4894$&$12$&$11.905$\\\hline
\end{tabular}
\caption{Higher order perturbative predictions for the reduced part of the 
semileptonic $\tau$ decay width using $a_\tau^{(3)}=0.2535$}
\label{tab1}
\end{center}
\end{table}
If the resummation does catch a dominant behavior then next several
orders of perturbation theory do not improve 
much and do not show wild asymptotic
behavior either. It forms a kind of dead zone where 
no qualitatively new behavior starts.

Results for the moments are presented in Table \ref{tab2}.
The label ``PT'' 
stands for finite order perturbation theory results \cite{ptest},
while ``$\rho_2$'' and ``$\rho_3$'' columns contain resummed results in
third and fourth order approximation for the $\beta$ function
\cite{adler}.
``Exp'' gives experimental number after direct integration of 
experimental data taken from the compilation
\cite{Eidelman}.
Last line indicates main source of the uncertainties
which are due to truncation of the perturbation theory series, 
an input 
value of $r_\tau^{exp}$ and systematic errors of the data.
\begin{table}\begin{center}
\begin{tabular}{|c|c|c|c|c|c|}\hline
&PT& $\rho_2$& $\rho_3$&exp\\\hline  
$R_0$&$2.28\pm 0.05$&$2.334$&$2.338$&$2.15$\\\hline
 $R_1$&$2.14\pm0.08$&$2.219$&$2.216$&$2.06$\\\hline
 $R_2$&$2.12\pm0.12$&$2.228$&$2.230$&$2.00$\\\hline
er-&trunca- &$\sim0.07$&$\sim0.07$&$\sim 10\%$\\
ror&tion & input& input&sys\\\hline
\end{tabular}
\caption{Predictions for moments of $e^+e^-$ annihilation rate 
through $\tau$ lepton width}
\label{tab2}
\end{center}\end{table}

Note that though we take into account the running of the 
coupling constant that is sensitive to an IR domain, due to a
particular way of
analytic continuation there is no direct problem of infrared
renormalons
\cite{ZakhAkh}.

At a given order of the $\beta$ function theoretical predictions with
resummation contain no error but due to input parameters ($r_\tau$ in
this case). Errors of summation method itself cannot be strictly given
however. As an estimate we use the change of predictions when passing
from one order to the next one, i.e.
$\rho_2 \rightarrow \rho_3$.
To find $\rho_3$ one needs the numerical value of $k_3$.
For the popular estimate of $k_3$ \cite{LeDiberderPich,KatSta}
one has $\rho_3=1.36$.
After the analysis done in \cite{adler}
we propose a slightly different number 
$\rho_3=2.0\pm 0.5$ that we have used for our predictions
(recall that in our normalization $\rho_1=0.79$, $\rho_2 =1.035$).

The only quantity that is given at the finite 
order of perturbation theory in our approach
is the $\beta$ function and one has to check its validity at 
least heuristically in terms of decreasing with order.
The worst pattern of convergence 
for the $\beta$ function in the course of the analysis
is given by (numerically)
\[
\beta(\alpha)\sim 1+0.284
+0.134+0.0467 \rho_3+\ldots
\]
at the point $\alpha=0.36$.
In all other points on the contour the convergence is better.
One can see that the numerical value of  $\rho_3$ is rather important
for application of discussed technique even though the change of
results is small. 

\section{CONCLUSIONS}
As conclusions to my talk I summarize results of investigation done
in \cite{resum,ptest,adler}:
\begin{itemize}
\item 
the coupling constant extracted from different
processes even in the same renormalization scheme 
(for instance, $\overline{\rm MS}$)
requires for precise comparison
an explicit mentioning of the resummation procedure
\item 
the recipe based on analytic continuation is rather stable against 
inclusion of higher order corrections to the $\beta$ function
\item 
moments of $e^+e^-$ annihilation at $\mu\sim m_\tau$ 
are estimated to be smaller than
required by pQCD with resummation of effects of running of the
coupling constant. 
\end{itemize}

\vskip 1cm
\noindent{\bf Acknowledgment}
\vskip 0.3cm

\noindent
This work is supported in part
by Russian Fund for Basic Research under contracts Nos. 96-01-01860
and 97-02-17065.

\end{document}